# Evaluation of the optical axis tilt of Zinc oxide films via noncollinear second harmonic generation.

F.A.Bovino,[1] M.C.Larciprete,[2,1*] A. Belardini,[2] C. Sibilia[2]

[1] *Quantum Optics Lab. , Elsag-Datamat Via Puccini 2 Genova, ITALY.*

[2] *Dipartimento di Energetica – Università di Roma "La Sapienza"*

*Via A. Scarpa 16, 00161 Roma, ITALY.*

## Abstract

We investigated noncollinear second harmonic generation form Zinc oxide films, grown on glass substrates by dual ion beam sputtering technique. At a fixed incidence angle, the generated signal is investigated by scanning the polarization state of both fundamental beams. We show that the map of the generated signal as a function of polarization states of both pump beams, together with the analytical curves, allows to retrieve the orientation of the optical axis and, eventually, its angular tilt, with respect to the surface normal.

[1] Corresponding author: mariacristina.larciprete@uniroma1.it

Second harmonic generation (SHG) processes are strongly dependent on the crystalline structure of the material thus offers the possibility to estimate the degree of structural order for samples produced by means of a certain deposition technique, and also to retrieve information about the orientation of the optical axis. The use of two pump beams, i.e. a noncollinear scheme, was firstly developed by Muenchausen [1] and Provencher [2]. More importantly, both Figliozzi [3] and Cattaneo [4] have shown that this technique allows the bulk and surface responses to be distinguished. In addition, Cattaneo has also demonstrated that the technique is very useful in surface and thin-film characterization [5-6].

We carried out noncollinear SHG measurements from ZnO films grown by dual ion beam sputtering. At a fixed incidence angle, the polarization state of both fundamental beams was systematically varied thus addressing all the different non-zero components of the nonlinear optical tensor [7-8]. The generated signal can be represented as function of polarization states of both pump beams. The result is a *polarization map* whose pattern is characteristic of the investigated crystalline structure. We here show that this method, which doesn't require sample rotation, is an useful tool to put in evidence the tilt angle of the optical axis of a nonlinear optical film with respect to the surface normal, for any material whose symmetry class implies an orientation of the optical axis almost perpendicular to the surface.

Zinc Oxide was chosen for the large energy gap value ($E_g$ = 3.37 eV) [9] and high nonlinear optical coefficients, of both second and third order, it offers. Second order nonlinear optical response has been shown in ZnO films grown by different techniques implying both high deposition temperature, as reactive sputtering [10] spray pyrolysis [11] laser ablation [12], and low deposition temperature, as laser deposition [2], and dual ion beam sputtering [13]. Generally, the reduced deposition temperature results in polycrystalline films, where the average orientation of crystalline grains, along with the resulting optical axis, can be tilted with respect to the ideal crystal, i.e. normal to sample surface.

ZnO crystalline structure, i.e. wurtzite, belongs to the noncentrosymmetric point group symmetry 6mm with a hexagonal primary cell. If the optical axis is normal to the sample surface, and correspond to the direction of the z axis, the second order susceptibility tensor, $\tilde{d}$, displays three nonvanishing components [14], along with the piezoelectric contraction, $d_{15}=d_{24}$, $d_{31}=d_{32}$ and $d_{33}$. Assuming Kleinmann symmetry [15] they reduce to two independent coefficients, $d_{15}=d_{24}=d_{31}=d_{32}$ and $d_{33}$. Furthermore, for an ideal wurtzite structure the nonzero elements are related to each other via the $d_{33} = -2 \cdot d_{31}$ [16]. If, on the other hand, the optical axis is tiled, with respect to the surface normal, a rotation must be applied to the nonlinear optical tensor, which is equivalent to the introduction of other nonvanishing terms in $\tilde{d}$.

Zinc oxide films, 400 nm thick, were deposited by means of a dual ion beam sputtering system onto 1 mm thick silica substrates [9]. X-ray diffraction (XRD) profiles performed on the obtained films show a peak in the θ/2θ curves located at $2\theta_2 \sim 34.4$ deg, originated from the 0002 planes of ZnO, as reported in Figure 1, indicating that the films are polycrystalline with the c-axis preferentially oriented about the surface normal [17].

SHG measurements were carried out by means of a noncollinear scheme working in transmission. The output of a mode-locked Ti:Sapphire laser system tuned at λ=830 nm (76 MHz repetition rate, 130 fs pulse width) was split into two beams of comparable power, while the temporal overlap of the incident pulses was controlled with an external delay line. The polarization of both beams was varied with two identical rotating half wave plates, which were carefully checked not to give nonlinear contribution since the two collimating lenses, 150 mm focal length, were placed thereafter. The sample was placed onto a motorized stage, allowing the variation of the sample rotation angle, α. The pump beams, laying in the same plane of incidence, were sent to intersect in the focus region with an angle of $\beta$= 9° and $\gamma$= -9°, measured with respect to $\alpha$ = 0° (see the insets in Figure 2). Thus, for a given $\alpha \neq$ 0°, the corresponding incidence angles of the two pump beams result to be $\alpha_1=\alpha-\beta$ and $\alpha_2=\alpha-\gamma$, respectively.

The interaction of two incident beams linearly polarized, tuned at $\omega_1$ and $\omega_2$, with a noncentrosymmetric material, produces a nonlinear polarization oscillating at the frequency $\omega_1+\omega_2$. Given the wave vectors' conservation law, $\vec{k}_{\omega 1} + \vec{k}_{\omega 2} = \vec{k}_{\omega 1+\omega 2}$, the generated beam is emitted nearly along the bisector of the aperture angle between the two pump beams. This beam was collected with an objective and focused on to a monomodal optical fiber coupled with a photon counting detector. Here, a set of optical low pass filters was used to further suppress any residual light at $\omega_1$ and $\omega_2$, while an analyzer allowed to select the desired SH polarization state.

The analytical expression of the effective susceptibility, $d_{\text{eff}}$ [18] depends on the $\tilde{d}$ tensor components, the polarization state of the three electric fields as well as on the fundamental beams incidence angles, $\alpha_1$ and $\alpha_2$. However, it can be significantly simplified for particular crystalline symmetries. Specifically, for the ZnO crystalline structure, considering four combination of polarization states of the two pump beams, $\hat{p}_{\omega 1} - \hat{p}_{\omega 2}$, $\hat{s}_{\omega 1} - \hat{s}_{\omega 2}$, $\hat{p}_{\omega 1} - \hat{s}_{\omega 2}$ and $\hat{s}_{\omega 1} - \hat{p}_{\omega 2}$, four different expressions for $d_{\text{eff}}$ are allowed, depending on the SH polarization state, i.e. either $\hat{P}$ or $\hat{S}$:

$$\begin{aligned}
d_{\text{eff}}^{pp \to P} &= -\cos(\alpha_{2\omega})d_{24}\left[\cos(\alpha_1')\sin(\alpha_2') + \cos(\alpha_2')\sin(\alpha_1')\right] + \\
&\quad + \sin(\alpha_{2\omega})\left[-d_{32}\cos(\alpha_1')\cos(\alpha_2') - d_{33}\sin(\alpha_2')\sin(\alpha_1')\right] \\
d_{\text{eff}}^{ss \to P} &= -\sin(\alpha_{2\omega})d_{31} \\
d_{\text{eff}}^{ps \to S} &= -d_{15}\sin(\alpha_1') \\
d_{\text{eff}}^{sp \to S} &= -d_{15}\sin(\alpha_2')
\end{aligned} \qquad (1)$$

where $\alpha'_1, \alpha'_2$ are the internal propagation angles of the two pump beams inside the sample, and $\alpha_{2\omega}$ is the angle of emission of the SHG inside the crystal. For pump beams linearly polarized with two generic polarization angles, $\phi_1$ and $\phi_2$, the $d_{eff}$ can be still calculated as $d_{eff} = \tilde{d}\vec{E}_1(\phi_1,\alpha_1):\vec{E}_2(\phi_2,\alpha_2)$.

The experimental measurements were obtained by rotating the two half-wave plates, in the range -180°-+180° for pump beam 2 and 0°-180° for pump beam 1. Different α as well as polarization state of generated beam were investigated, although we here report only the measurements of $\hat{S}$ polarized SH signal for the sake of brevity.

When the analyzer set to $\hat{S}$-polarization, i.e. $\phi$=90° degrees, the maxima of SH signal occur when the two pump beams have crossed polarization. The last two equations in (1) show that this condition is not symmetrical for positive and negative rotation angles α, since the position of the two beams, differently polarized, with respect to the sample surface is not symmetrical. When $\alpha$=35° (Figure 2a), the absolute maxima take place when pump 1 is $\hat{s}$-polarized and pump 2 is $\hat{p}$-polarized, i.e. $\phi_1$ equal to ±90° and $\phi_2$ to either 0° or 180°, while relative maxima occur in the reverse situation, that is pump 1 is $\hat{p}$-polarized and pump 2 $\hat{s}$-polarized. According with the theoretical model, we found the opposite behaviour when the rotation angle $\alpha$ is set to -35°, as shown in Figure 2b, thus the absolute and relative maxima appear to be inverted with respect to the previous situation. For both α values, when the two pumps are equally polarized, either $\hat{s}$ or $\hat{p}$, the nonlinear optical tensor do not allow $\hat{S}$-polarized SH signal.

Experimental plots obtained for $\alpha$ = 9° are shown in Figure 2c. In this particular condition, the absolute maxima still require $\hat{s}$-polarization for pump 1 and $\hat{p}$-polarization for pump 2 (i.e. $\phi_1$= ± 90° and $\phi_2$ equal to either 0° or 180°), whereas the relative maxima totally disappeared. This peculiarity can be reasonably explained considering that $\alpha$ = 9° corresponds to a situation such that the pump beam

1 is normally incident onto the sample, as can be seen form the inset. For an anisotropic uniaxial crystal with its optical axis perpendicular to sample surface, a normally incident wave always experiences the ordinary refractive index, whatever its polarization angle, thus pump beam 1 always corresponds to an $\hat{s}$-polarized beam. As a consequence for $\alpha = 9°$, the condition to get a relative maximum, i.e. $\hat{p}$-polarization for pump1 and $\hat{s}$-polarization for pump 2, is never fulfilled, being replaced with the condition equivalent to two pumps both having $\hat{s}$-polarization. This combination do not allow $\hat{S}$-polarized SH signal, thus the relative maxima disappear. Moreover, we found out that the experimental configuration where one of the pump beams is normally incident onto the sample, is particularly sensitive to the orientation of the optical axis.

The experimental curves were fully reconstructed using the expression for the effective second order optical nonlinearity in noncollinear scheme, assuming the Kleinmann symmetry rules. Dispersion of both the ordinary and extraordinary refractive indices of ZnO, $n(\lambda)$ was taken from reference 3. We show in Figure 3a the calculated curve for $\alpha = 9°$, when the optical axis is assumed to be perpendicular to sample surface. If compared with the theoretical one, the experimental curve appears to be shifted towards higher $\phi_2$. This difference between experimental and theoretical curves suggest that the optical axis may be averagely tilted with respect to the surface normal. This is a reasonable assumption, taking into the low temperature deposition technique which was employed.

An angular tilt of the optical axis was then introduced in the analytical model through a rotation matrix, applied on the $d_{eff}$. The rotation produces the arising of new terms in the nonlinear optical tensor. In Figure 3b we show the new curve, calculated for a tilt of 2° around the x-axis, as shown in the inset. The obtained theoretical curve displays the same $\phi_2$-shift evidenced in the experimental curves, thus indicating that the film has a partially oriented polycrystalline structure, as shown by the X-ray analysis, but the orientation of the optical axis is not exactly normal to the film surface. Similar curves were calculated by tilting the optical axis, along with the nonlinear optical tensor, around the other two reference axes. For the investigated crystalline symmetry group, 6mm, a rotation about the z-axis do not produce any change in the nonlinear optical tensor. On the other side, a rotation about the y-axis produce an analogous shift in the $\hat{S}$-polarized SH pattern, but also a modification in the $\hat{P}$-polarized SH pattern which was not compatible with the corresponding experimental curves.

In conclusion, we investigated second order nonlinear optical properties of ZnO films deposited by dual ion beam sputtering, with a noncollinear experimental setup. We show that, from the contour plot of the generated signal as a function of polarization states of both pump beams, important information on the crystalline structure of the films can be deduced. The polarization scanning method adopted is a valid and sensitive tool to probe the orientation of the optical axis and to evidence possible angular tilt with respect to surface normal.

## Acknowledgements.

M.Centini is kindly acknowledged for helpful discussion and interesting comments. ZnO films were grown in collaboration with F.Sarto at the Division of Advanced Physics Technologies of ENEA (Roma, Italy).

**Captions to Figures.**

**Figure 1.** X-ray diffraction profiles of the ZnO films, 400 nm tchik, grown by dual ion beam sputtering onto 1mm glass substrate.

**Figure 2.** Second harmonic signal as a function of the polarization angle of the first pump beam ($\phi_1$) and the second pump beam ($\phi_2$). Sample rotation angle was fixed to (a) $\alpha = 35°$, (b) $\alpha = -35°$ and (c) $\alpha = 9°$, respectively. Polarization state of the analyzer is set to $\hat{S}$, i.e. $\phi = 90°$. The insets show the condition for the polarization states of the two pump beams to get the absolute maxima.

**Figure 3.** Theoretically calculated curves of $\hat{S}$-polarized second harmonic signal as a function of the polarization angle of the first pump beam ($\phi_1$) and the second pump beam ($\phi_2$), calculated for the optical axis (a) normal to the sample surface and (b) tilted about the x-axis of 2 degrees. Sample rotation angle is $\alpha = 9°$. Thick arrow in the insets represents the optical axis orientation.

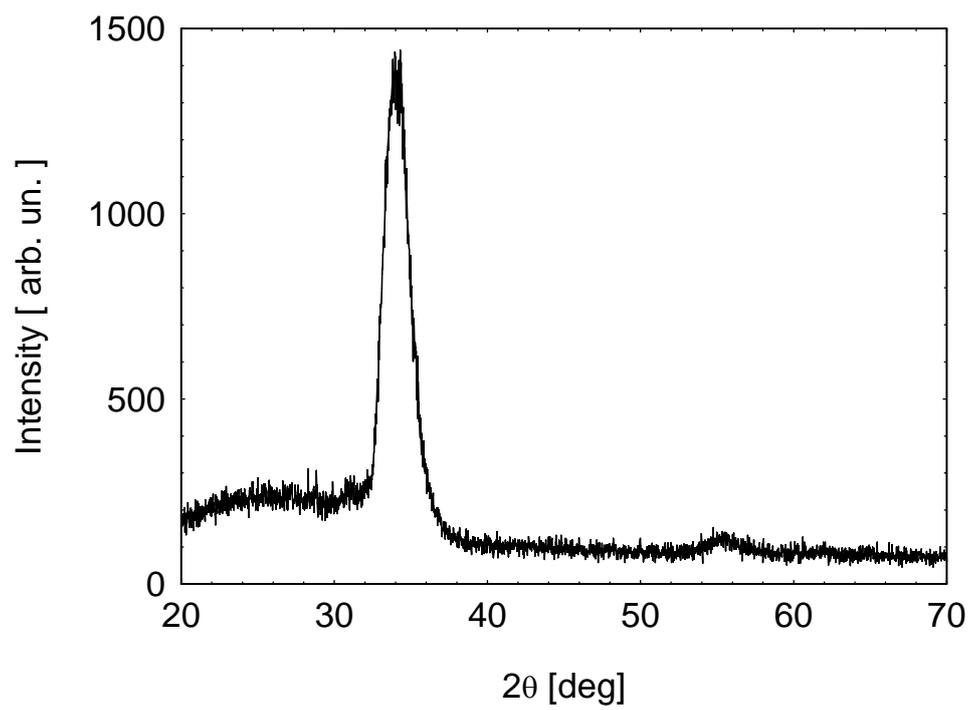

**Figure 1**

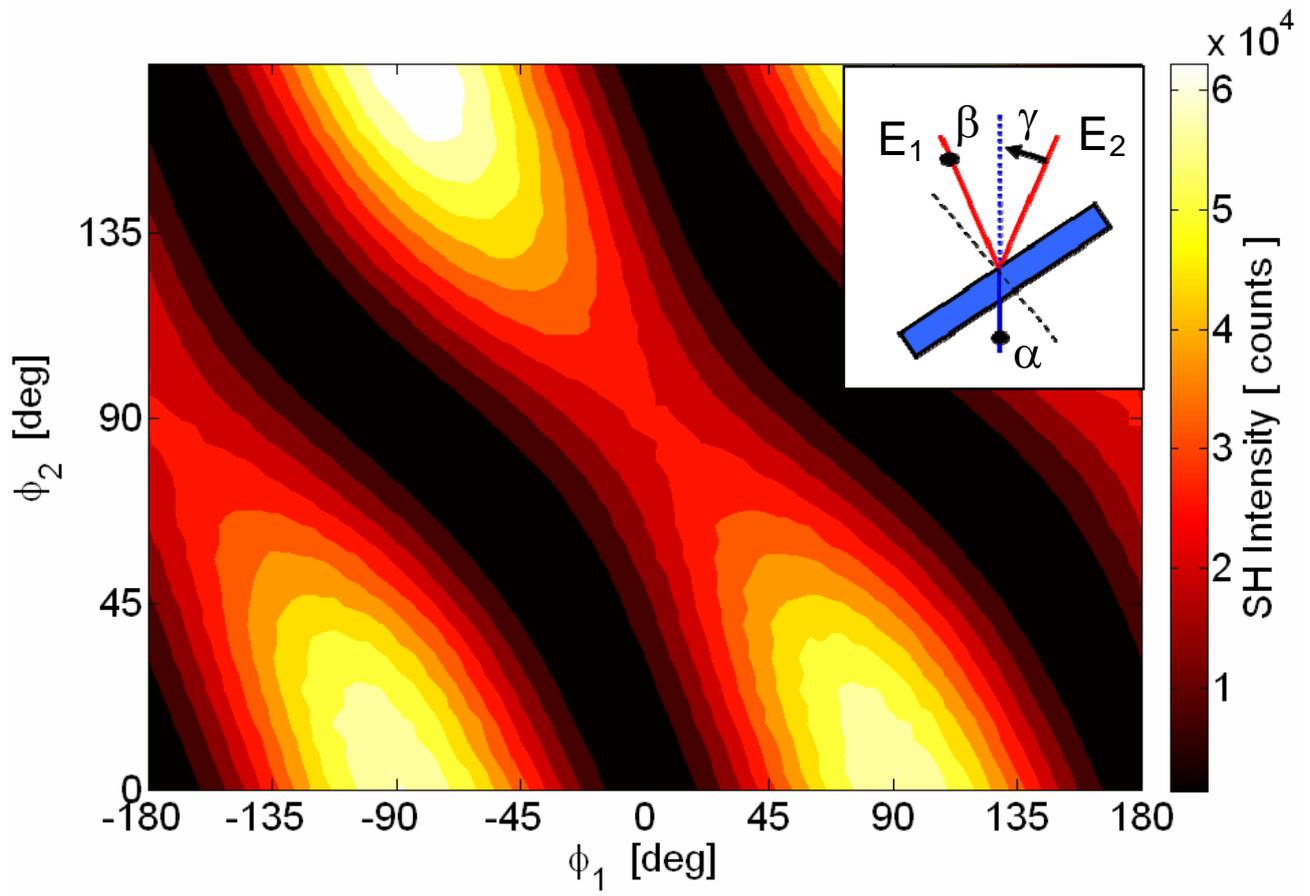

**Figure 2 (a).**

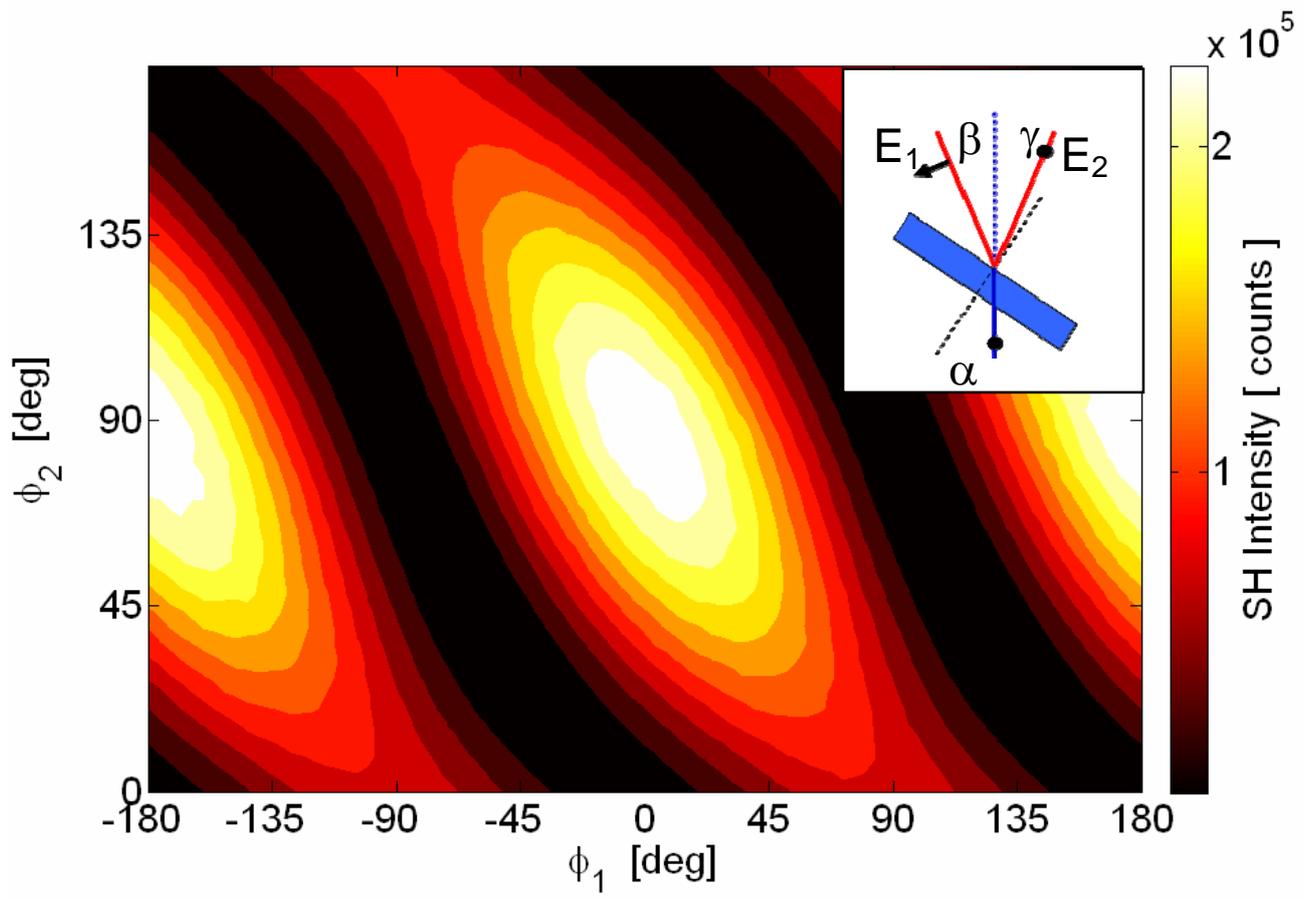

**Figure 2 (b).**

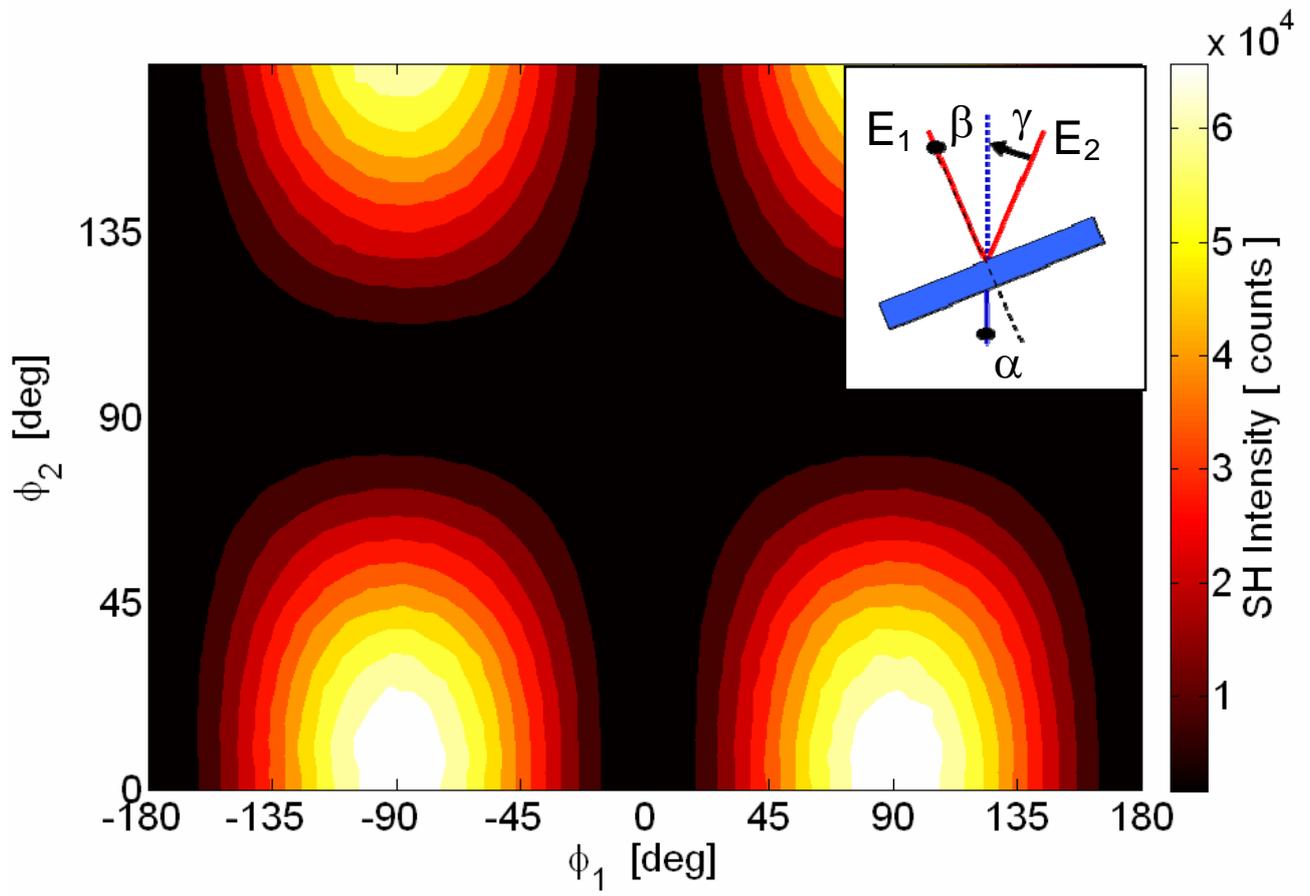

**Figure 2 (c).**

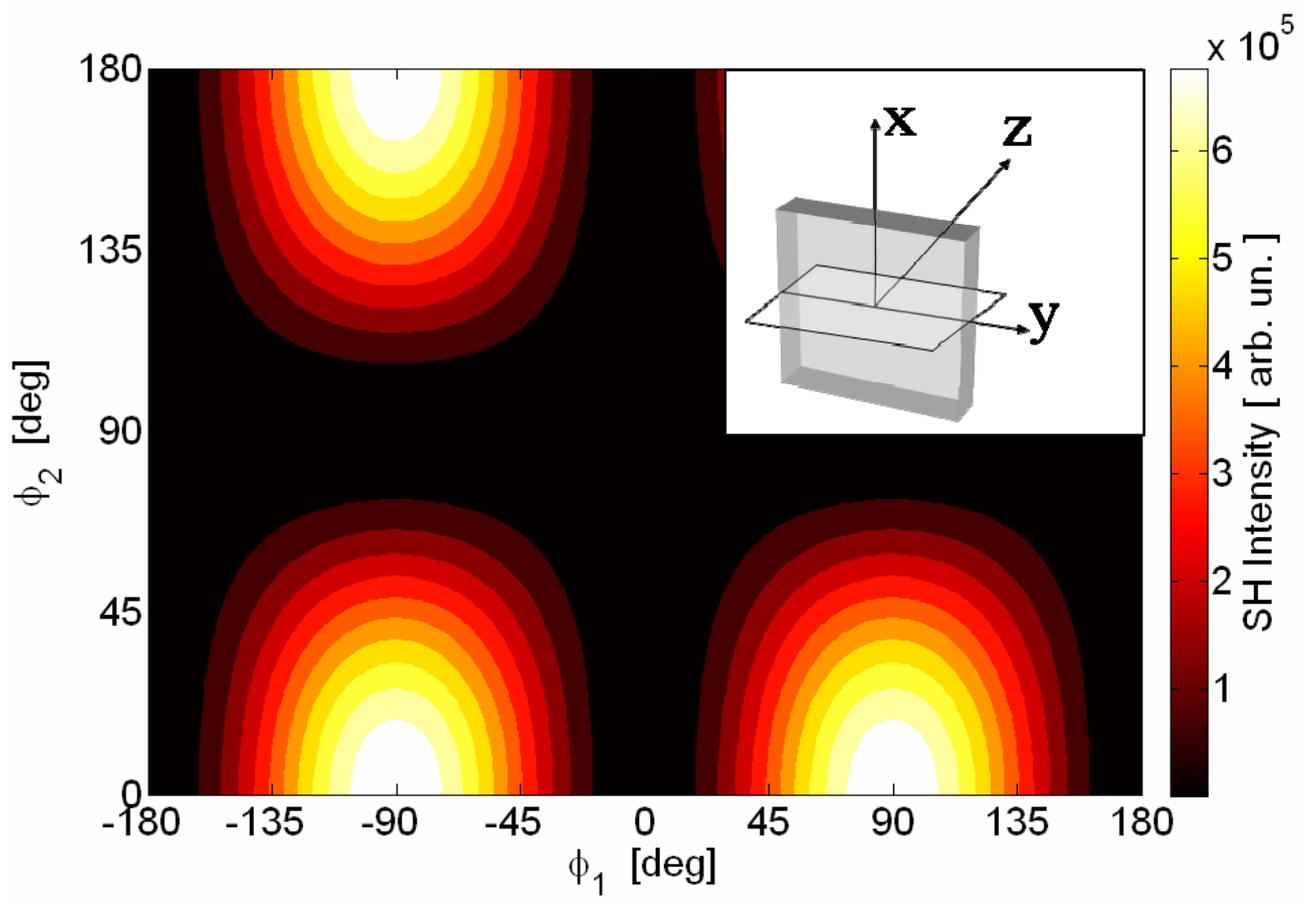

**Figure 3 (a).**

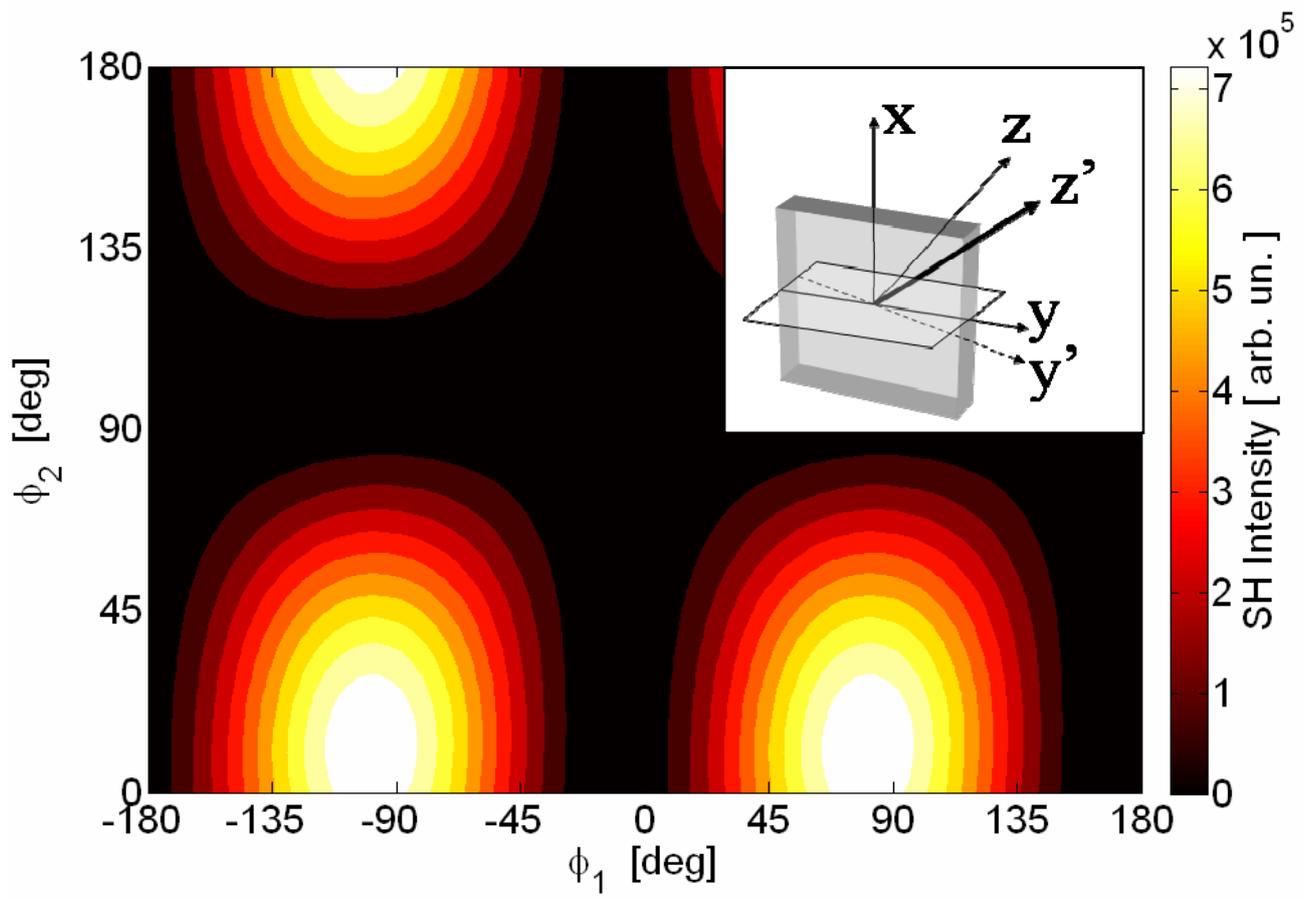

**Figure 3 (b).**

**List of References:**

1   R.E. Muenchausen, R.A. Keller and N.S. Nogar, J.Opt.Soc.Am.B. **4**, 237-241 (1987).

2   P. Provencher, C.Y. Côté and M.M. Denariez-Roberge, Can. J. Phys. **71**, 66-69 (1993).

3   P. Figliozzi, L. Sun, Y. Jiang, N. Matlis, B. Mattern, M.C. Downer, S.P. Withrow, C.W. White, W.L. Mochán and B.S. Mendoza, Phys. Rev. Lett. 94, 047401/1-047401/4 (2005).

4   S. Cattaneo and M. Kauranen, Phys. Rev. B 72, 033412/1-033412/4 (2005).

5   S.Cattaneo and M.Kauranen, Opt. Lett. 28, 1445-1447 (2003).

6   S. Cattaneo, E. Vuorimaa, H.Lemmetyinen, M. Kauranen, J. Chem. Phys. 120, 9245-9251 (2004).

7   J. Jerphagnon and S.K. Kurtz, J. Appl. Phys. **41**, 1667-1681 (1970).

8   D.Faccio, V.Pruneri and P.G.Kazansky, Optics Letters, 25, pp.1376-1378 (2000).

9   R. Blachnik, J. Chu, R. R. Galazka, J. Geurts, J. Gutowski, B. Hönerlage, D. Hofmann, J. Kossut, R. Lévy, P. Michler, U. Neukirch, T. Story, D. Strauch, A. Waag, "Zinc oxide (ZnO)" in *Landolt-Börnstein-Group III Condensed Matter, Semiconductors: II-VI and I-VII Compounds; Semimagnetic Compounds*, 41B, 52-53, Springer-Verlag (1999).

10  G.Wang, G.T.Kiehne, G.K.Wong, J.B.Ketterson, X.Liu and R.P.H.Chang, Appl. Phys. Lett. 80, 401-403 (2002).

11  U.Neumann, R.Grunwaid, U.Griebner, G.Steinmeyer and W.Seeber, Appl. Phys. Lett. 84, 170-172 (2004).

12  H.Cao, J.Y.Wu, H.C.Ong, J.Y.Dai and R.P.H.Chang, Appl. Phys. Lett. **73**, 572-574 (1998).

13  M.C.Larciprete, D.Passeri, F.Michelotti, S.Paoloni, C.Sibilia, M.Bertolotti, A.Belardini, F.Sarto, F.Somma and S.Lo Mastro, J. Appl. Phys **97**, 23501-23506 (2005).

14  A.Yariv, *Optical Electronics 4th ed.* (Saunders College Publishing, U.S.A.,1991).

[15] V.G. Dmitriev, G.G. Gurzadyan and D.N. Nikogosyan, *Handbook of nonlinear optical crystals* (Springer, Berlin, 1997).

[16] B.F.Levine, Phys.Rev.B 55, pp.13630 (1997).

[17] K.S.Weiβenrieder and J.Muller, Thin Solid Films **300,** 30-41 (1997).

[18] P.A.Franken and J.F.Ward, Rew. Modern Physics **35**, 23-39 (1963).